\documentclass[journal=jacsat,manuscript=article]{achemso}

\usepackage[version=3]{mhchem} 
\usepackage[T1]{fontenc}       
\usepackage{graphicx}
\usepackage{epstopdf}
\usepackage{color}

\author{Hirofumi Yanagisawa}
\affiliation[ETH1]
{Institute for Quantum Electronics, ETH Z\"{u}rich, CH-8093 Z\"{u}rich, Switzerland}
\alsoaffiliation[MPQ]
{Max Planck Institute for Quantum Optics, D-85748 Garching, Germany}
\alsoaffiliation[Postech1]
{Max Planck POSTECH/KOREA Res. Init., Pohang, 37673, South Korea }
\email{hirofumi.yanagisawa@mpq.mpg.de}

\author{Vahur Zadin}
\affiliation[estonia1]
{Intelligent Materials and Systems Lab, Institute of Technology, University of Tartu, Nooruse 1, 50411 Tartu, Estonia}
\alsoaffiliation[Helsinki]
{Helsinki Institute of Physics and Department of Physics, P.O. Box 43, 00014 University of Helsinki, Finland}

\author{Karsten Kunze}
\affiliation[ScoopM]
{{\footnotesize Scientific Center for Optical and Electron Microscopy, ETH Z\"{u}rich, CH-8093 Z\"{u}rich, Switzerland}}

\author{Christian Hafner}
\affiliation[ETH2]
{Laboratory for Electromagnetic Fields and Microwave Electronics, CH-8092 Z\"{u}rich, Switzerland}

\author{Alvo Aabloo}
\affiliation[estonia1]
{Intelligent Materials and Systems Lab, Institute of Technology, University of Tartu, Nooruse 1, 50411 Tartu, Estonia}

\author{Dong Eon Kim}
\affiliation[Postech2]
{Department of Physics, POSTECH, Pohang, 37673, South Korea}
\alsoaffiliation[Postech1]
{Max Planck POSTECH/KOREA Res. Init., Pohang, 37673, South Korea }

\author{Matthias F. Kling}
\affiliation[MPQ]
{Max Planck Institute for Quantum Optics, D-85748 Garching, Germany}
\alsoaffiliation[LMU]
{Physics Department, Ludwig-Maximilians-Universit\"{a}t M\"{u}nchen, D-85748 Garching, Germany.}

\author{Djurabekova Flyura}
\affiliation[Helsinki]
{Helsinki Institute of Physics and Department of Physics, P.O. Box 43, 00014 University of Helsinki, Finland}

\author{J\"{u}rg Osterwalder}
\affiliation[University of Zurich]
{Physik Institut, Universit\"{a}t Z\"{u}rich, Winterthurerstrasse 190, CH-8057 Z\"{u}rich, Switzerland}

\author{Walter Wuensch}
\affiliation[CERN]
{European Organization for Nuclear Research, CERN, 1211 Geneva 23, Switzerland}

\title[nanoletters]
  {Laser-induced asymmetric faceting and\\ growth of nano-protrusion on a tungsten tip}

\abbreviations{}
\keywords{}

\begin{document}

\begin{abstract}

Irradiation of a sharp tungsten tip by a femtosecond laser and exposed to a strong DC electric field led to gradual and reproducible surface modifications. By a combination of field emission microscopy and scanning electron microscopy, we observed asymmetric surface faceting with sub-ten nanometer high steps. The presence of well pronounced faceted features mainly on the laser-exposed side implies that the surface modification was driven by a laser-induced transient temperature rise -- on a scale of a couple of picoseconds -- in the tungsten tip apex. Moreover, we identified the formation of a nano-tip a few nanometers high located at one of the corners of a faceted plateau. The results of simulations emulating the experimental conditions, are consistent with the experimental observations. The presented conditions can be used as a new method to fabricate nano-tips of few nm height, which can be used in coherent electron pulses generation. Besides the direct practical application, the results also provide insight into the microscopic mechanisms of light-matter interaction. The apparent growth mechanism of the features may also help to explain the origin of enhanced electron field emission, which leads to vacuum arcs, in high electric-field devices such as radio-frequency particle accelerators.

\end{abstract}


A metallic nano-tip with sharpness of a few nanometers can provide very bright and spatially coherent electron waves \cite{fu01, nagaoka01, binh92, fink86, oshima02}. This is highly beneficial in many applications such as electron diffraction, microscopy, and holography \cite{Zuo03, wang00,tonomura05}. During irradiation by laser pulses a nano-tip is also expected to generate pulsed electron waves with high brightness and coherence \cite{Lee73,hommelhoff06a,yanagisawa09,yanagisawa11,hommelhoff11}. Such electron pulses enable experiments with high temporal and spatial resolution for investigations of ultrafast phenomena in solids \cite{ropers14}. A higher-performance pulsed electron source could make a significant contribution to future investigations of the dynamics in ultrafast devices \cite{schiffrin13}. 

So far various well-established methods have been used to fabricate nano-tips \cite{fu01, nagaoka01, binh92, fink86, fujita07, fujita08, rokuta11}. Typically, a nano-tip is grown on a larger metallic tip by applying strong DC fields \cite{nagaoka01, binh92, fink86, fujita07, fujita08, rokuta11}, heating \cite{binh92, fujita07, fujita08, rokuta11} or depositing metal \cite{fu01, fink86}. Fabrication of a nano-tip {\it in situ} for laser-induced electron pulse applications is, however, not a simple task because of the presence of optical elements used in the laser focusing system which must remain very clean. Thus formation of a nanotip on a larger tip apex by the laser itself will significantly simplify and advance the fabrication process.

In addition, the understanding of microscopic mechanisms governing metal surface evolution under strong electric fields is crucial for other much larger scale applications. Identification and study of atomic scale surface dynamics is practically impossible on metal surfaces of centimeter dimensions. Strongly enhanced field emission currents often are measured from macroscopic surfaces with a sub-micron flatness. Fitting field emission data from such systems to the Fowler-Nordheim equation \cite{wuensch14} systematically requires introducing a so-called field enhancement factor, $\beta$, which is typically between 50 and 200, see for example Ref. [\cite{wuensch14}]. The $\beta$ factor is associated with a local increase of the applied electric field, which is needed to be able to explain measured currents. In relatively low field devices this can be attributed to contaminants, but the field enhancement factors of 50 to 200 persist, for example, in high-gradient accelerating structures which have very clean, highly processed surfaces and surface electric fields above 200 MV/m \cite{wuensch14, grudiev14,chen12,degiovanni16}. A theoretical explanation of the $\beta$ factor suggest formation of nanoprotrusions \cite{djurabekova13}, but efforts to identifying the features which cause the field enhancement through electron microscopy have been unsuccessful, given the inherent difficulty of identifying nano-scale features on centimeter-sized surfaces and correlating with field emission measured centimeters away. The nano-tip system appears to be an excellent means to study field emission enhancing features in detail given the limited size of the tip. The features of the type described in this paper may turn out to be a dominant source of enhanced field emission sites.

In this Letter we report the observation of a nanotip formed by femtosecond laser pulses on the apex of a tungsten tip held under a strong DC electric field. We investigated the surface modification induced by the laser and the applied field by using field emission microscopy (FEM) and high-resolution scanning electron microscopy (SEM). Facets surrounded by sub-ten nm high steps appeared asymmetrically, i.e., mainly on the laser-exposed side. We detected a nano-tip grown at one of the corners of a step edge after approximately 5 hours of near-continuous pulsed laser irradiation and a strong DC electric field. Theoretical simulations of local field enhancements near a metal tip apex due to different surface nanofeatures were  consistent with the experimental observations. This is the first direct experimental observation of a nanotip grown under laser irradiation from an initially atomically smooth metal surface.

\begin{figure}
\begin{center}
\includegraphics[scale=0.18]{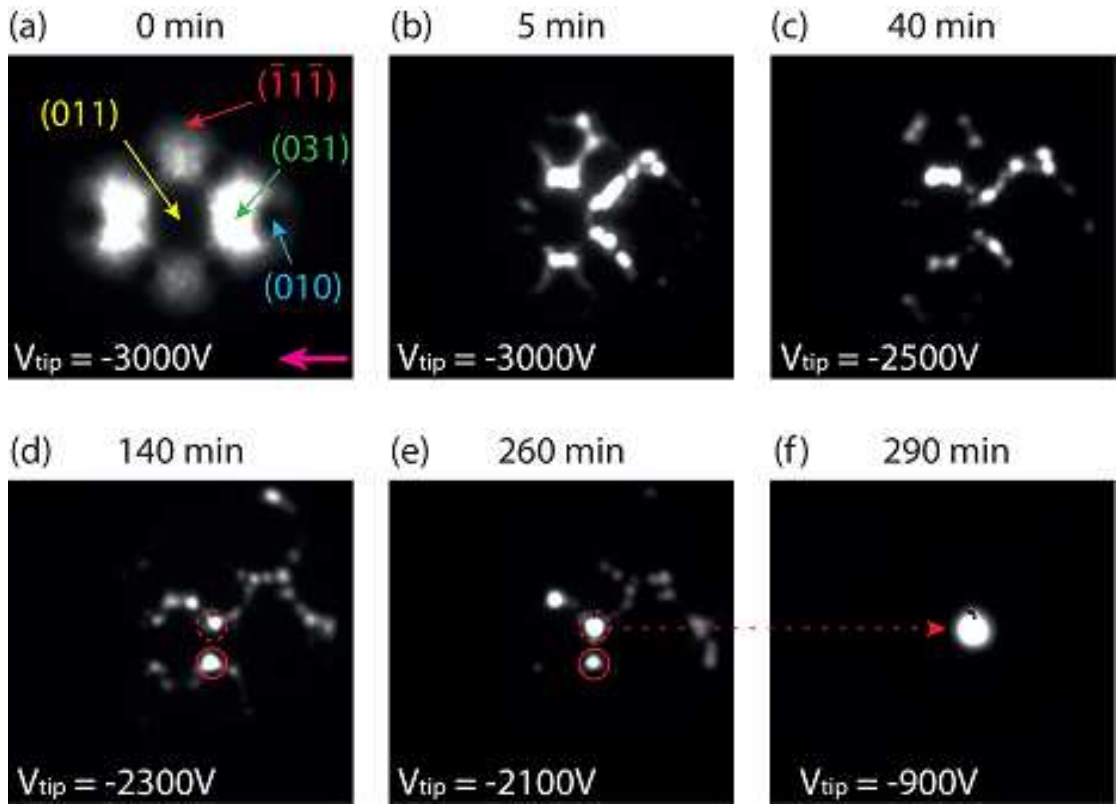}
\end{center}
  \caption{ (a) - (f) Field emission patterns without laser some time after laser irradiation. During the laser irradiation, $V_{tip}$ was -3000V and $P_{L}$ was 180 mW. The propagation direction of the laser is indicated by a pink arrown in (a). Tip voltages and accumulated irradiation time are shown on each corresponding image. (a) is the pattern from the tip apex cleaned by flash heating prior to irradiation. Here, the surface indices of some facets are also indicated.}\label{fig:FEM}
\end{figure}

\textsf{{\bf Evolution of field emission pattern under laser irradiation.}} Applying high voltages on a metallic tip with nanometer sharpness resulted in a field concentration only at the tip apex and the strong field drives field emission (FE) via electron tunneling through the surface barrier \cite{gomer93}. The FE current exponentially depends on work function and local fields on the tip apex \cite{gomer93}. Thereby FEM provides geometrical information on the tip apex, and can achieve a spatial resolution of 1 or 2 nm. 

In our experiments we used a single crystal tungsten tip with the tip apex orientated in the [011] direction. A clean tip apex can be obtained by heating the tip \cite{gomer93, sato80}. With increasing heating temperature, the radius of curvature becomes bigger; the radius of curvature was approximately 250 nm in this study. When the tip apex is clean, FEM images show a characteristic pattern such as the one in Fig. \ref{fig:FEM}a. Details about the FEM experiment can be found in the Supplementary Information. In this pattern, the intensity distribution roughly represents the work function map because the local-field distribution changes moderately over the tip apex. Here the most intense emission spots seen in Fig. \ref{fig:FEM}a correspond to the (310) crystal faces with the lowest work function. On the other hand, an increase of field emission currents can be due to local field enhancement caused by geometric nanofeatures form on the tip apex. The appearance of laser-induced nanofeatures described in this Letter was recorded via this mechanism.

The cleaned tip was subject to pulsed laser irradiation with a laser power of 180 mW. During the laser irradiation, a DC voltage of -3000 V was applied to the tip. Periodically, we paused the laser irradiation and recorded FEM images. The exposure time for the FEM images was 2.5 seconds. The tip voltages were set in each case to provide sufficiently strong electron signals on the FEM fluorescent screen, but avoiding saturation of the signal. Fig. \ref{fig:FEM} shows the evolution of the FE patterns which we observed during the experiment for a laser irradiation time which was approximately 5 hours. The time displayed in each figure is the accumulated laser irradiation time. As one can see, already after the first 5 minutes of laser irradiation, the  emission pattern had changed significantly from the original one. The extended bright areas in the original FE pattern from the faces with the low work function were replaced  by dotted-line features, which appeared asymmetrically, mainly on the laser-exposed side. During laser irradiation, the vacuum in the chamber was maintained at a pressure below $3 \cdot 10^{-10}$ mbar. It is worth noting that the FEM images did not change significantly when the tip was left in such vacuum for 5 minutes without laser irradiation (see Supplementary Information). Such a change might have occured if the enhance emission was due to absorption of residual-gas-related contaminants. In addition, a DC electric field alone does not produce an asymmetric pattern during FEM experiments \cite{nagaoka01, gerstl09}. Therefore, it is apparent that the observed change in the FE pattern was induced by the laser. As will be discussed later, these features form as a consequence of the step edges of faceted surfaces where local fields are enhanced.

The dotted-line features remained in the FE pattern for more than 4 hours of the laser irradiation (see the image at 260 minutes in Fig. \ref{fig:FEM}e). However the necessary applied tip voltage for imaging was gradually decreased with irradiation time. This implies an increase of field enhancement due to formation of smaller nano-structures \cite{fujita07, fujita08}. After the first two hours of laser irradiation, we noticed well pronounced spots of intense electron emission, see Figs. \ref{fig:FEM}d and \ref{fig:FEM}e. In these figures, two particularly bright spots are marked by red circles. Their relative intensities change from image to image. After almost 5 hours of laser irradiation (290 minutes), the brightest spot in Fig. \ref{fig:FEM}e eventually took over the rest of the spots and is visible as a single field emitting spot in Fig. \ref{fig:FEM}f. For this condition the tip voltage had to be significantly lowered, to maintain the level of image quality and avoid saturation of the detector. The appearance of the bright spot and the strong reduction of the tip voltage indicate the formation of a nano-tip \cite{fujita07, fujita08}. 
 
We emphasize that the laser-induced changes in FE patterns are reproducible. By heating the tip, the tip apex was reshaped to its original clean condition producing the expected FE pattern on the FEM fluorescent screen, same as in Fig. \ref{fig:FEM}a. Repeating the same procedure with the laser irradiation resulted in a similar evolution of the FEM images as shown in Fig. \ref{fig:FEM}. We reproduced nano-tips in the described fashion twice on different tips.


\begin{figure}
\begin{center}
\includegraphics[scale=0.15]{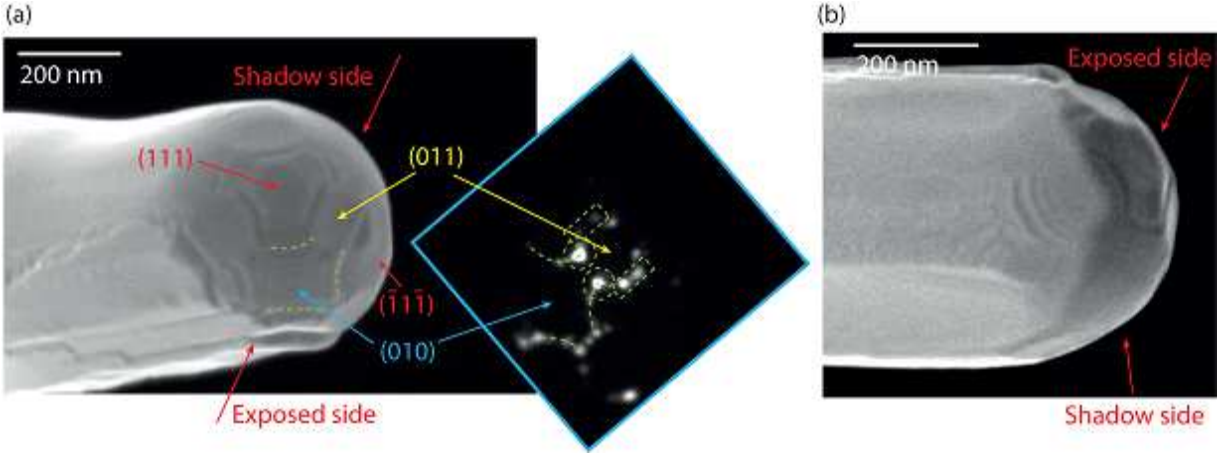}
\end{center}
  \caption{(a) and (b) SEM images of the tungsten tip apex with a nano-tip from different view points, acquired in high resolution (immersion) mode using secondary electron detector (through-the-lens detecotr) at 2kV and working distance of 4mm. The white scale bars indicating 200 nm. The inset of (a) is the electron emission pattern identical to Fig. \ref{fig:FEM}(d). }\label{fig:tip}
\end{figure}
  
In order to investigate the origin of the dotted-line structures and the bright spot, we removed the sample from the vacuum chamber and inspected the tip apex with a high-resolution SEM (FEI Magellan) \cite{SEMref}. Fig. \ref{fig:tip}a and \ref{fig:tip}b show SEM images of the tip apex from two different view points, the laser-exposed and the shadow sides. From these two viewing perspectives the  asymmetric surface restructuring on the tip apex can be seen. Faceting was mainly observed on the exposed side, including the shank, as shown in Fig. \ref{fig:tip}a. The surface on the shadow side, by contrast, remained rather smooth as it is shown in Fig. \ref{fig:tip}b. Note that some laser-induced surface modifications on tip apexes were observed in previous work \cite{gerstl09,koelling11,sha08}, but not the gradual growth of prominent features, which we observed in our experiments, have been reported so far.

SEM and FEM images are compared in the inset of Fig. \ref{fig:tip}a. They reveal curved features on the exposed side of the tip apex in the SEM image which are similar to the pattern in the FEM image (highlighted by dashed green lines). From the similarity in symmetry and shape, we deduce that the electrons were emitted from these curved features outlining the facet with the (010) crystal face. Other faces can be identified based on the symmetries and relative position with respect to the (010) surface as shown in Fig. \ref{fig:tip}a. 

\begin{figure}
\begin{center}
\includegraphics[scale=0.12]{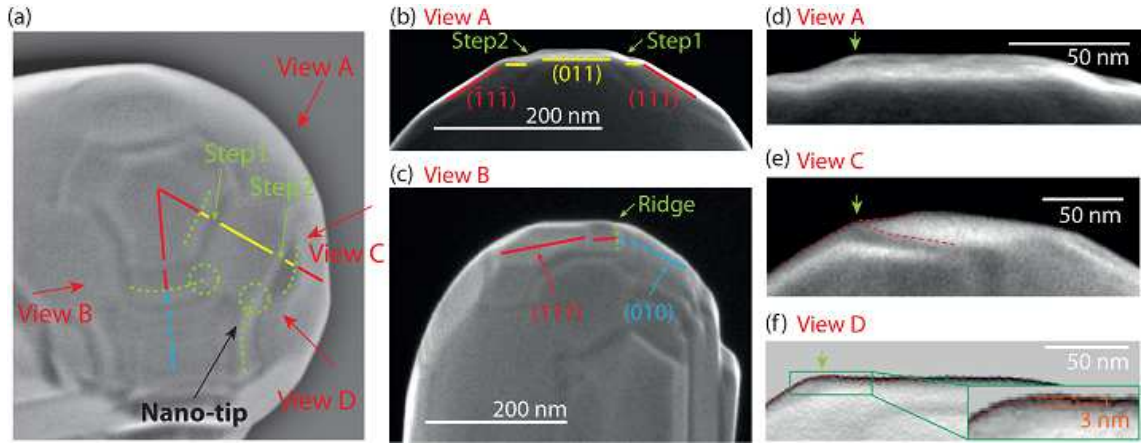}
\end{center}
\caption{(a)-(f) SEM images of the tungsten tip apex from different view points and for different magnifications. In all images, the yellow, red and blue lines indicate (011), (111) and (010) type facets, respectively. The dashed lines and circles indicate ridges and corners, respectively, formed by intersection of facets. (a) shows only the apex of the tip from the same image as in Fig. \ref{fig:tip}a). Edges of structures observed in (a) and (f) are intensified for more clear visualization. The different view points of imaging the tip apex and shown in (b)-(f) are roughly indicated as Views A-D. The white scale bars are 200nm in (b) and (c), and 50nm in (d)-(f). The green arrow indicate the location of the nano-tip in (d)-(f). }\label{fig:SEM}
\end{figure}

To clarify the origin of the surface features observed in the SEM images, the latter were taken from several view points, View A-D, indicated in Fig. \ref{fig:SEM}a. The line structures, especially those indicated in Fig. \ref{fig:SEM} as Step1 and Step2 , were clearly resolved as steps with a height of approximately 8 nm in the side-view (View A in Fig. \ref{fig:SEM}b). In the same image, the facets of (011) and (111) orientations are also clearly observed, for convenience these are indicated by yellow and red lines, respectively. The angles between yellow and red lines are approximately 35 degrees, which agrees with the theoretical value of the angle between (011) and (111) crystallographic planes in the tungsten lattice. The available view of the (010) surface makes it more difficult to determine if there are steps, however it is likely that there are, in analogy with those clearly seen on the (011) surface. Indeed there is some indication that there may even be more steps, and candidates are marked by blue lines in Fig. \ref{fig:SEM}c.


The facets which have been formed intersect with each other and to form ridges as indicated by a dashed line in Fig. \ref{fig:SEM}a and \ref{fig:SEM}c. In addition, we can clearly identify the corners formed by the intersection of three facets (011), (111) and (010). These corners are marked by green dashed circles in Fig. \ref{fig:SEM}a. The ridges and corners have sharper edges so that the corresponding local fields are  higher than at other places on the tip apex. Indeed, the spots seen in the FEM images which form curved lines around (010) in the inset of Fig. \ref{fig:tip}a appear to lie on the ridges surrounding the (010) facet. The other field emission spots, marked by rectangles and circles in the inset, can be consistently explained by ridges and corners, respectively. Now most importantly, a nano-tip appears to have grown at one of the corners (the position of the tip is indicated by a black arrow in Fig. \ref{fig:SEM}a). This is observed in the FEM images as the bright spot which appears near the end of the exposure time. In addition, we have observed a faint feature of a nano-protrusion protruding from the corner, indicated by the green arrow in Figs. \ref{fig:SEM}d-\ref{fig:SEM}f. In order to emphasize the feature, the red dashed lines are drawn to show outlines in Fig. \ref{fig:SEM}e and \ref{fig:SEM}f. The height of the observed nanotip is estimated to be approximately 3 nm.


\begin{figure}
\begin{center}
\includegraphics[scale=0.13]{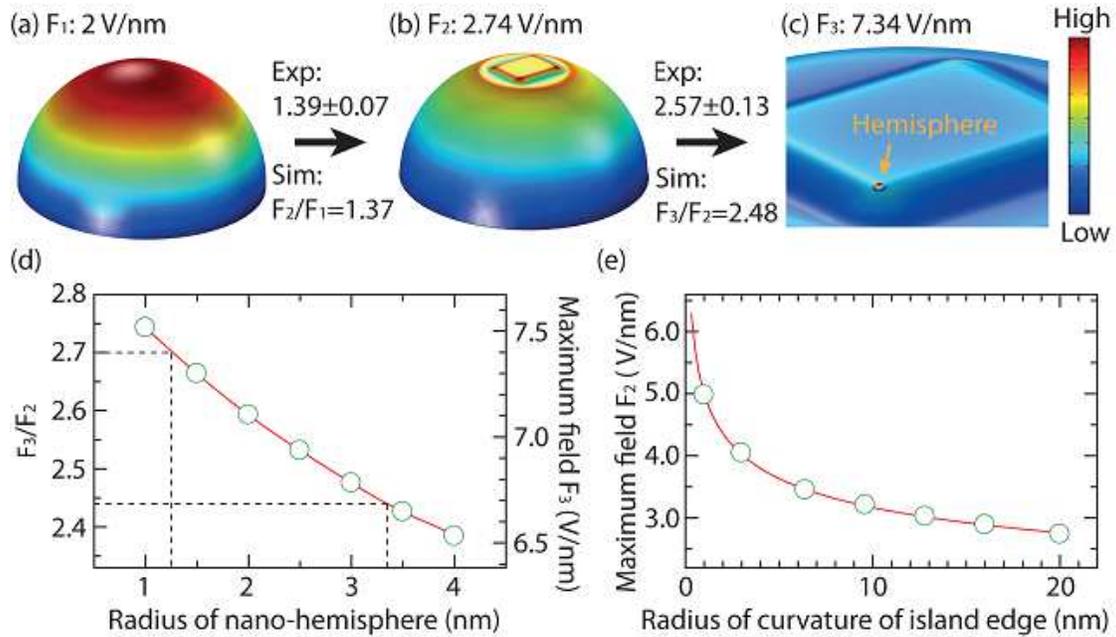}
\end{center}
  \caption{(a) - (c) Local field distributions on a tip apex with different shapes. (a) Hemispherical tip apex, (b) hemispherical tip with a rectangular plateau on top of the apex and (c) nano-hemisphere at one corner of the rectangular plateau of (b). See the text for details. Maximum local fields for the three models, $F_{1}$, $F_{2}$ and $F_{3}$, are shown in each figure. (d) Changes of maximum fields with varying radius of the nano-hemisphere. The maximum fields are shown on right side axis. The left side axis shows changes of field enhancement factors with respect to that of (b). (e) Changes of maximum fields with varying radius of curvature of the plateau in (b).}\label{fig:sims}
\end{figure}

\textsf{{\bf Local field enhancement and simulations of a nanotip.}} We have investigated through simulation the field enhancements of the step and tip features described above in order to verify consistancy between features observed in SEM and FEM images. The changes of the enhancement factors in experiments can be  approximately estimated from the threshold voltages where the FE current reaches an observable level by the two-dimensional electron detector in our experimental setup (see Supplementary Information). The threshold voltages for the sample conditions in Fig. \ref{fig:FEM}a, \ref{fig:FEM}e and \ref{fig:FEM}f were 2500 V, 1800 V and 700V, respectively. Therefore, field enhancement factors change by factor of 1.39 from the sample conditions of Fig.\ref{fig:FEM}a to \ref{fig:FEM}e and by factor of 2.57 from \ref{fig:FEM}e to \ref{fig:FEM}f. The errors of these values are approximately 5\%, which is determined from the resolution of the voltage measurements and the intensity resolution of the screen.

We have performed the simulations of local field distributions and determined geometrical field enhancement factors for three different geometries. The geometries were set to match the above-mentioned experimental observations. The  details of the simulations can be found in the Supplementary Information. Three geometries can be described as follows. The first (Fig. \ref{fig:sims}a) is a perfect hemispherical shape with a radius of curvature of 250 nm to represent the as-prepared tip apex. The second has a rectangular plateau on the tip apex as shown in Fig. \ref{fig:sims}b. The plateau is surrounded by the steps and sharp corners to represent the (011) facet seen in Fig. \ref{fig:SEM}b. We modeled this structure since the the field emission current in the experiment was the strongest at one of the corners of such a plateau (cf. Fig. \ref{fig:tip}a) and the dimensions of the rectangular plateau in the model were the same as in the SEM observation in Fig. \ref{fig:SEM}d and \ref{fig:SEM}f. The third is shown in Fig. \ref{fig:sims}c and has a nano-hemisphere with the radius of 3 nm (see Fig. \ref{fig:SEM}f) placed at one of the corners of the plateau, where the local fields caused by the plateau are at a maximum, as shown in Fig. \ref{fig:sims}b.

In the simulations the applied field was the same for all three geometries, and was chosen to result in the maximum field $F_{1} = 2$ V/nm for the hemispherical tip as shown in Fig. \ref{fig:sims}a. This is a typical field strength which results in field emission from a tungsten tip. Comparing now the values of the maximal local fields in all three geometries, we observe a gradual enhancement of the local fields very similar to that measured in the experiment. Indeed, the value of the maximal local field from Fig. \ref{fig:sims}a ($F_{1}$) to \ref{fig:sims}b ($F_{2}$) changed by $F_{2}/F_{1}=1.37$ (compared to 1.39 measured in the experiment) and from Fig. \ref{fig:sims}b ($F_{2}$) to \ref{fig:sims}c ($F_{3}$) by factor $F_{2}/F_{1}=2.48$ (compared to 2.57 in the experiment). 

The observed size of the nano-tip is also supported by further field distribution simulations. By varying the radius of the nano-hemisphere, we analyzed the evolution of the factor $F_{3}/F_{2}$. As can be seen in Fig. \ref{fig:sims}d, the enhancement of the local field meets the experimental value $2.57 \pm 0.13 $ if the radius of the nano-hemisphere is within the range from 1 nm to 3.5 nm. (the dashed lines in Fig. \ref{fig:sims}d are a guide for the eye). This result is consistent with our SEM observation within its experimental errors. 

A pyramidal shape of a surface feature obtained by an intersection of three facets was also simulated to check if such a feature alone could cause the observed field enhancement. This could be due to a strong local enhancement due to the atomic level sharpness of the top of the pyramid ~\cite{fujita07, fujita08, rokuta11}. As can be seen in Fig. \ref{fig:sims}e, the simulations did not confirm this hypothesis. Sharpening the pyramidal top appeared to be insufficient to account for the observed field enhancement. Even a curvature radius of 0.3 nm (interatomic distance of the tungsten lattice) of the simulated plateau corner did not result in a sufficiently high local field. Hence all the observations and simulations coherently support the presence of a nano-tip at a corner of the surface facets with the well pronounced surrounding steps.


\textsf{{\bf Heating effect of femtosecond pulsed laser irradiation.}} It is known that applying a strong DC bias on a heated tip drives faceting on the tip apex \cite{fujita07, fujita08, rokuta11}. Although we do not heat our sample during the experiment, the laser pulses irradiating the tip apex cause local heating, which in combination with the strong electric field, leads to the surface modification. Heating of the crystal lattice occurs through energy absorption from the laser pulses by electrons and the subsequent energy transfer from the excited electrons to the lattice. Although the elevated lattice temperature lasts up to nanoseconds after each laser pulse, the  temperature rises quite high during the first 10 picosends after laser excitation. More importantly, within a few picoseconds after laser excitation, the temperature distribution is very asymmetric with higher temperatures on the laser-exposed side \cite{kasevich12}. After the first period of a hot and asymmetric temperature distribution, the temperature significantly decreases and the distribution becomes uniform. Therefore, the observed asymmetric faceting on the exposed side must be driven by the transient thermal effects on a time scale of a couple of picoseconds. A simple estimation of the lattice temperature results in the value of 4200 K (see Supplementary material), which can be considered as the maximum possible lattice temperature in the studied regime. This temperature is sufficient to melt the surface layer of tungsten, at least, for a short period of time, which is a likely explanation to liquid-like behavior of surface tungsten atoms. The temperature evolution due to laser pulses and surface atom behavior are subject to further studies and will be reported in following publications. For the nano-tip, it appears that its growth requires the stronger electric fields caused by the facet edge because of its location there. Hence, the laser-induced faceting is the essential initial step which then results in the growth of a nano-protrusion. 


In summary, we have observed asymmetric faceting and growth of a nano-protrusion on a tungsten tip apex induced by a combination of sub-ten femtosecond laser pulses and strong DC fields. This is a new method to create a nano-tip and can likely be also used for processing nano-patterns on the sub-ten nanometer scale by faceting. The growth of nano-tips under intense fields is also of potential importance to the particle accelerator community. A strong rf-field may induce nano-tip growth on the surfaces of rf cavities in the same way as on the tip. Vacuum arcing, initiated from enhance field emission, limits the strength of the rf-fields and thus the achievable accelerating gradient~\cite{wuensch14}. Therefore, investigation of the formation of nano-tips under intense electric fields may be important input for finding technological solutions for an increasing gradient in normal conducting radio frequency accelerators. It is, moreover, very difficult to observe a tip of a few nanometers size grown on a macroscopic surface of the cavity surface. Hence the laser-induced surface modification on the tip presented in this work has the potential to become an excellent research tool for investigating and understanding the growth mechanism of surface nano-features under extreme conditions.

\begin{acknowledgement}
This work was supported by the Swiss National Science Foundation through the {\it Ambizione} (grant number PZ00P2\_131701) and the {\it NCCR MUST}, Kazato Research Foundation, Estonian Science Foundation grant PUT 57, the Academy of Finland (grant number 296969), the National Research Foundation of Korea through Global Research Laboratory Program [Grant No 2009-00439] and Max Planck POSTECH/KOREA Research Initiative Program [Grant No 2011-0031558], the DFG via SPP1840 and the EU via the grant ATTOCO. We thank Prof. C. Oshima and Dr. Lukas Gallmann for fruitful discussions and M. Baer for technical support.
\end{acknowledgement}

\begin{suppinfo}

\begin{figure}
\begin{center}
\includegraphics[scale=0.17]{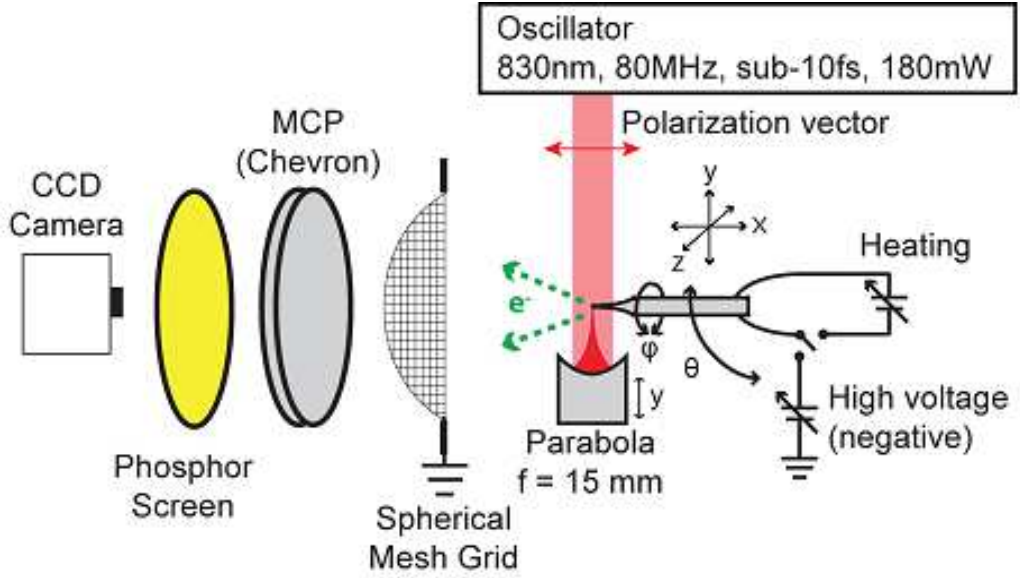}
\end{center}
  \caption{Schematic diagram of the experimental setup. A tungsten tip is mounted inside a vacuum chamber. Sub-10fs laser pulses are generated outside the vacuum chamber. A parabolic mirror is installed next to the tip to focus the laser onto the tip apex. The polarization vector of the laser is parallel to the tip axis. The two-dimensional detector was used to observe the electron emission patterns from the tip.}\label{fig:setup}
\end{figure}

\textsf{{\bf Experimental setup.}}The FEM experiments were performed by using an experimental setup as shown schematically in Fig. \ref{fig:setup}. A tungsten tip is mounted inside a vacuum chamber (base pressure: $1 \cdot 10^{-10}$ mbar). Laser pulses are generated in a Ti:sapphire oscillator (center wavelength: 830 nm; repetition rate: 80 MHz; pulse width: sub-10 fs; laser power: 180 mW) and introduced into the vacuum chamber. A parabolic lens (focal length: $f$ = 15 mm) is located just next to the tip to focus the laser onto the tip apex; the diameter of the focused beam is approximately 3.5 $\mu$m ($1/e^2$ radius). Linearly polarized laser light was used with the polarization vector parallel to the tip axis. In order to position the tungsten tip apex to the focus of the laser, the tip holder can be moved along three linear axes ($x$, $y$, $z$) and two rotational axes for azimuthal ($\varphi$, around the tip axis) and polar ($\theta$, around the $z$ axis) angles by using piezo stages. The tip can be negatively biased for FE. A phosphor screen with a Chevron-type double-channelplate amplifier in front of the tip is used to record the emission patterns. All the measurements were done at room temperature.

\textsf{{\bf Changes of FEM images with time.}} Changes of FEM images with time under ultra high vacuum condition (pressure: $3.6 \cdot 10^{-10}$ mbar) are shown in Fig. \ref{fig:contami}. The images were recorded by a position sensitive detector with an experimental setup explained elsewhere \cite{yanagisawa09}. Up to 30min, the FEM images stay similar. The behavior is significantly different from changes in Fig. 2 in the main text.

\begin{figure}
\begin{center}
\includegraphics[scale=0.16]{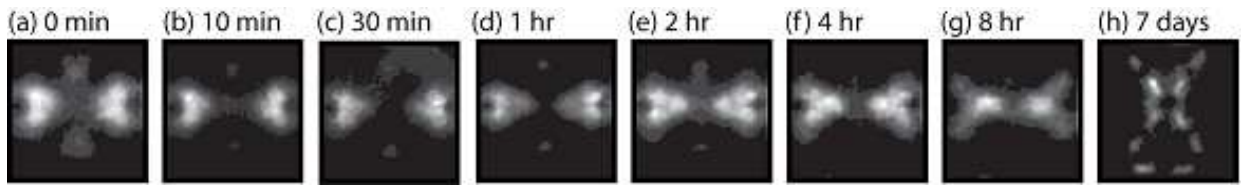}
\end{center}
  \caption{(a)-(g) changes of FEM images with time under ultra high vacuum condition. }
  \label{fig:contami}
\end{figure}

\textsf{{\bf Details of local DC field simulations.}}
The simulations of local DC fields were performed by using Comsol Multiphysics \cite{comsol}. Assuming no space charge density in vacuum around the tip, the electric field distribution was obtained by solving the Laplace equation for the defined geometries \cite{zadin14}. Since the electric field calculations near the surfaces with small radii of curvature are sensitive to numerical errors, a special care was taken to ensure the validity of the results. A tetrahedral mesh with quadratic shape functions was used in all simulations, and mesh convergence was carefully checked. The numerical solution was obtained by using a conjugate gradient iterative solver with algebraic multigrid preconditioner, the relative tolerance of the calculations was set to $10^{-9}$.

\textsf{{\bf Estimation of transient lattice temperature due to a laser pulse.}} The maximum lattice temperature rise in the studied regime was estimated by dividing the absorbed energy density from a laser pulse at the tip surface by the lattice heat capacity. The absorbed energy density for our laser parameters where the asymmetric surface modifications were observed was estimated by using the method described in Ref. [\cite{kasevich12}], giving 10 ${\rm {kJ}/{cm^{3}}}$. A tungsten lattice heat capacity of 2.6 ${\rm {J}/{Kcm^{3}}}$ was used, which is determined by Dulong-Petit law \cite{kittel}. The calculated transient temperature rise due to a single laser pulse was added to the initial room temperature of the tungsten tip (300 K) giving a maximal temperature of 4200 K.

\end{suppinfo}

\end{document}